\documentclass[sigplan, 10pt]{acmart}

\acmConference[MIDDLEWARE'19]{ACM/IFIP/USENIX MIDDLEWARE Conference}{December 09--13, 2019}{Davis, CA, USA}
\acmYear{2019}
\acmISBN{} 
\acmDOI{} 
\startPage{1}

\setcopyright{rightsretained}

\usepackage[subtle]{savetrees}

\usepackage{microtype}
\usepackage{graphicx}
\usepackage{subfigure}
\usepackage{booktabs} 
\usepackage{hyperref}
\usepackage{xspace}
\usepackage{color,soul}
\usepackage{url}
\newcommand{\IBM}{IBM\xspace}

\newcommand{\DLaaS}{FfDL\xspace}

\newcommand{\shep}[1]{#1}

\begin{document}

\title{\DLaaS : A Flexible Multi-tenant Deep Learning Platform}


\author{K. R. Jayaram, Vinod Muthusamy, Parijat Dube, Vatche Ishakian, Chen Wang, Benjamin Herta, Scott Boag, Diana Arroyo, Asser Tantawi, Archit Verma, Falk Pollok, Rania Khalaf}
\affiliation{
  \institution{IBM Research}            
  \city{Yorktown Heights, NY and Cambridge, MA}
  \country{USA}                    
}

\begin{abstract}

Deep learning (DL) is becoming increasingly popular in several application domains and has made several new application features
involving computer vision, speech recognition and synthesis, self-driving automobiles, drug design, etc. feasible and accurate. 
As a result, large scale ``on-premise'' and ``cloud-hosted'' deep learning platforms have become essential infrastructure 
in many organizations. These systems  accept, schedule, manage and execute DL training jobs at scale.  

This paper describes the design, implementation and our experiences with \DLaaS, a DL platform 
used at \IBM.
We describe how our design balances dependability with scalability, elasticity, flexibility and efficiency. 
We examine \DLaaS qualitatively through a retrospective look at the lessons learned from 
building, operating, and supporting \DLaaS; and quantitatively through a detailed empirical evaluation of \DLaaS, 
including the overheads introduced by the platform for various DL models, the load and performance 
observed in a real case study using \DLaaS\ within our organization, the frequency of various faults observed 
including faults that we did not anticipate, and experiments demonstrating the benefits of various scheduling policies.
\DLaaS\ has been open-sourced.

\end{abstract}







\maketitle

\section{Motivation and Introduction}~\label{sec:introduction}

Deep learning (DL), a form of machine learning (ML), has been applied to fields including computer vision, speech recognition, natural language processing, social network filtering, machine translation and bioinformatics where it has produced results comparable to or even superior to human experts~\cite{nature}. A confluence of several factors has given rise to the popularity of deep learning:

\begin{itemize}
	
	\item In several scenarios, deep neural networks (NN) are able to learn features in an unsupervised manner. 
	Traditionally, speech recognition, image classification, etc. involved handcrafting features for particular applications. With DL, feature engineering can often be avoided. 
	
	\item Distributed feature representation over several neurons. The NN not only learns how to learn these features, but also determines how they can be combined well. This distributed feature representation is therefore very powerful; having more degrees of freedom and the ability to approximate complex functions well compared to other ML techniques.
	
	\item New GPU technologies have increased the efficiency of large scale matrix computations, and enabled larger DL models to fit into GPU RAM.
	
	\item Advances in interconnection and data center networking technologies like NVLink, Infiniband and 100G Ethernet enable distributed DL training algorithms to transfer large amounts of training data and synchronize large model parameters without the network becoming a bottleneck.

	\item Open source DL frameworks such as TensorFlow, PyTorch, Caffe, Torch, Theano, and MXNet reduce the effort and skill set required to design, train, and use DL models.
	
\end{itemize}

\emph{Training deep neural networks}~\cite{nature} using large datasets (typically tens or hundreds of TB) is 
computationally expensive and time consuming (typically several days depending on type of hardware). 
It involves making several 
passes over the same data set, which often has to be streamed from external storage. Often, the same
training workload is run with different hyperparameters, to determine the optimal set of hyperparameters.
To execute DL training workloads reliably multiple times, data scientists at our organization had to 
build their own infrastructure stack. In addition to increasing ``stack sprawl'', this involved a steep learning 
curve because data scientists are often not experts in cluster management, failure detection and response,
taking consistent checkpoints, etc. Errors were common and achieving dependability required a lot of work 
which was duplicated across teams.

Even when teams were able to successfully set up their infrastructure, there was the issue of
\emph{low resource utilization}. While advances in hardware have enabled DL to scale, said hardware 
remains expensive and should be effectively utilized to obtain good returns on investment. A distributed deep 
learning \emph{platform} (DLP) helps organizations (like ours) utilize expensive hardware effectively, 
and enables all developers in the organization to share DL infrastructure. 
Such platforms also reduce the barrier to entry by enabling our  developers to focus on 
training neural nets and choosing hyper-parameters rather than focusing on installation, 
configuration and fault tolerance. A good DLP is also the foundation of ``deep learning as-a-Service''
offerings by cloud providers to other organizations and the general public.

The obvious question at this point is ``Why do we need a middleware platform specialized
for DL training workloads''?
Modern compute clusters are typically managed by \emph{container orchestration} 
middleware like Kubernetes~\cite{kubernetes} and 
Mesos~\cite{mesos} because datacenter applications are typically deployed as containers. 
It has been demonstrated that container orchestration middleware are able to handle
many resource sharing, scheduling and failure detection tasks for reasonably sized compute clusters
(hundreds of machines) as long as the application developer is willing and able to containerize their
application and choose the right middleware abstraction (i.e. the right Kubernetes or Mesos abstraction)
while deploying his application. However, they have
key shortcomings as far as managing DL training workloads:

\begin{itemize}

	\item While cluster managers do a great job abstracting away low-level hardware and application placement details,
	provide useful APIs and include tools to help simplify application deployment, they are still considered
	too ``low-level'' for use by data scientists. For example, in the case of Kubernetes, data scientists are typically not
	interested in understanding the difference between a ``Job'' and a ``Deployment'' or configuring network
	policies for communication between pods and their constituent containers. Instead, they want to focus on 
	writing training code in Python, picking optimal hyperparameters, specifying the location from where training data can be fetched,
	and where checkpoints and the trained model have to be stored. They expect the DL platform to figure out the rest.

	\item Some DL training workloads need specialized scheduling algorithms, specifically a form of \emph{gang scheduling} 
	which is not natively supported in many cluster managers.

	\item Cluster managers are able to provide generic job status updates (e.g., Pending, Running, Failed, Completed), but
	are unable to provide additional DL-specific job statuses (e.g., \shep{DOWNLOADING, PROCESSING, STORING, HALTED, RESUMED} etc.)

\end{itemize}

This paper describes \DLaaS, a dependable distributed deep-learning platform in the cloud 
used at \IBM. This paper makes three key contributions:

\begin{itemize}

\item We articulate the requirements of a cloud native distributed deep learning platform as distilled from customer use cases. We also outline challenges in meeting these requirements based on our experience building and operating a number of cloud services. Section~\ref{sec:challenges} describes these requirements and challenges and incorporates the evolution in our understanding as we developed the platform and worked with users.

\item The bulk of the paper presents the design of the \DLaaS cloud-based distributed DL platform. The platform serves as middleware between the deep learning frameworks and hardware resources. We devised techniques to support scalability and fault tolerance to frameworks designed with neither in mind, added custom scheduling algorithms as out-of-the box cluster manager schedulers are not tuned for DL workloads, and built storage drivers optimized for deep learning access patterns, all while meeting stringent cloud security requirements without sacrificing the performance of DL jobs. Section~\ref{sec:dlaas} describes the detailed architectural elements of \DLaaS and how the requirements and challenges in Section~\ref{sec:challenges} are addressed.

\item We present both qualitative and quantitative examinations of the \DLaaS platform.
Section~\ref{sec:lessons} presents a retrospective look at the lessons we learned building, operating, and supporting \DLaaS.
Then, Section~\ref{sec:evaluation} reports a detailed evaluation of \DLaaS, including the overheads introduced by the platform training a variety of deep learning models, the load and performance observed in a real case study using \DLaaS within our organization, the frequency of various faults observed including faults that we did not anticipate, and experiments demonstrating the benefits of custom schedulers.

\end{itemize}

The paper concludes with a description of relevant related work in Section~\ref{sec:related} and concluding remarks in Section~\ref{sec:conclusions}.

\section{\DLaaS : Design Goals and Challenges} \label{sec:challenges}

\paragraph{Design Goals:} To be effective, a distributed DL platform (DLP) should be \emph{easy to use, flexible, dependable, scalable, and efficient}. At a high level:

\begin{itemize}

	\item \emph{Ease of Use.} Ideally, a user has to focus only on his or her training application, and not 
worry about secondary issues such as setup, security and failure handling. Users should be able to start running their
DL training jobs at scale with minimal modifications, if any, within tens of minutes.
	
	\item \emph{Flexibility} usually means that the platform should support several frameworks. Like programming languages, DL developers, over time, tend to prefer developing in a specific framework (e.g., TensorFlow vs. Caffe).
	
	\item \emph{Dependability} means that the platform should be highly available, reliable, handle faults in a robust manner and also be secure and maintainable.

	\item \emph{Scalability}, in this case, refers to ability of the DL platform to manage both small and large workloads and 
	compute clusters with similar efficacy and quality of service. 	
	
	\item \emph{Efficiency} (of the platform) usually means that the overheads introduced by the platform to achieve the aforementioned goals (especially flexibility and dependability) and the response time of the platform to external requests must be minimal. We would like to note that this paper is about a DL platform, and not about the efficiency of a specific framework (e.g., Caffe vs PyTorch) or a distribution architecture (parameter servers vs. peer-to-peer). Hence, we define efficiency in terms of the additional overhead and response time introduced by the platform.
	
\end{itemize}

\paragraph{Challenges:} Users expect cloud services to be highly available and scalable, and 
there are best practices and architecture patterns around microservices that many services 
have employed to achieve dependability requirements. A DL service, 
however, differs from most services, as outlined below.

\begin{itemize}

	\item The technology stack, including DL frameworks and middleware support for GPUs, aren't 
	designed for a cloud environment or with dependability goals in mind.
        Distributed DL frameworks, such as Caffe2, Horovod~\cite{horovod} and PyTorch, assume stable dedicated hardware resources. 
	Process or hardware faults are not handled, and users are expected to write their code to take snapshots 
	and resume from them. This differs from microservices which are typically built to be \emph{loosely coupled and stateless}.

	\item DL training jobs consist of running user code. There is first of all a challenge in running any untrusted 
	code securely in a multi-tenant environment.  There is also the issue that it is not feasible to analyze the 
	user code to infer requirements (such as the number of training iterations), or instrument it to take 
	snapshots or resume from snapshots.
	
	\item DL jobs are long running, often taking several days over which hundreds of thousands of 
	iterations are processed. The consequences of failure can therefore be significant, potentially losing several days of work and thereby also wasting expensive hardware resources.

	\item DL jobs are GPU-heavy, and are engineered to exploit the massive SIMD parallelization in GPUs and maximize 
	GPU utilization. This increases heat generated by GPU servers in the datacenter, and server machine failures 
	(typically reboots, power downs, etc.) are not uncommon.  While container orchestration platforms such as Kubernetes have some support to allocate GPUs as a resource, they are unable to detect and handle many types of GPU failures. Also, GPU workloads cannot easily migrate.
	
	\item DL jobs impose a heavier load on datacenter networks. DL training algorithms make several passes over the data set, which can be tens (or sometimes hundreds) of TB. At these sizes, data cannot be stored locally and typically has to be streamed over the network (either from a cloud-based store or NFS) for \emph{each} pass. 
	
\end{itemize}

In addition, operating a flexible multi-framework, multi-tenant DL platform supporting 
single node and distributed DL jobs requires supporting the following features:

\begin{itemize}

    \item Deploying a DL job requires orchestrating the provisioning of a number of resources including GPUs, network connections, storage volumes, and credentials. The provisioning needs to be atomic, both to avoid wasted resources when provisioning only partially completes due to resource constraints, and to reliably reclaim resources after faulty provisioning steps.

    \item Given that DL jobs are long running, users expect periodic and accurate status updates, such as whether the job is DEPLOYING or PROCESSING. These status updates should be dependable because users use associated timestamps for job profiling and debugging. Further since the users are charged for their actual GPU usage, transparency about the true status of jobs is important. 

    \item Reliable streaming of logs from the job, irrespective of the stage it is in, even if it crashes/fails. This is key for users to debug their jobs.

    \item DL frameworks are so flexible that, for example, a TensorFlow job can contain arbitrary customer code. Hence, for multi-tenancy, DL jobs must be isolated from \DLaaS system processes, and from each other.

    \item Support for both user-directed and configurable automatic periodic checkpointing,
    given the longevity of DL jobs.

    \item Resilience to node and job crashes. Both failure detection and recovery are important because users expect to be notified when DL jobs are restarted, as the convergence of a model training can differ slightly when a job experiences a failure.

\end{itemize}

\section{\IBM \DLaaS : A Detailed Overview} \label{sec:dlaas}

\DLaaS is a cloud-hosted and multi-tenant dependable distributed DL platform used to train DL models at \IBM,
that has been open-sourced\footnote{https://github.com/IBM/FfDL}.  
We present an overview of \DLaaS and outline how it achieves high availability, scalability and isolation while supporting 
several popular DL frameworks.  \DLaaS is a cloud-native application architected as a set of loosely-coupled 
microservices communicating with each other using gRPC. Logically, \DLaaS has three layers: 
(1) the Core Services layer, consisting of the API, Lifecycle Manager (LCM) and Training Metrics microservices; 
(2) the Platform layer comprising the infrastructure that the core services rely on including Docker~\cite{docker}, 
Kubernetes~\cite{kubernetes}, etcd~\cite{etcd}, and MongoDB~\cite{mongodb}; and 
(3) Helpers, which are components that are spawned as part of the DL job during execution. 
Helpers perform failure detection, status recording and updates, log collection, data and results 
transfer, and metrics collection.

\begin{figure}[htb]
    \centering
    \includegraphics[width=0.5\columnwidth]{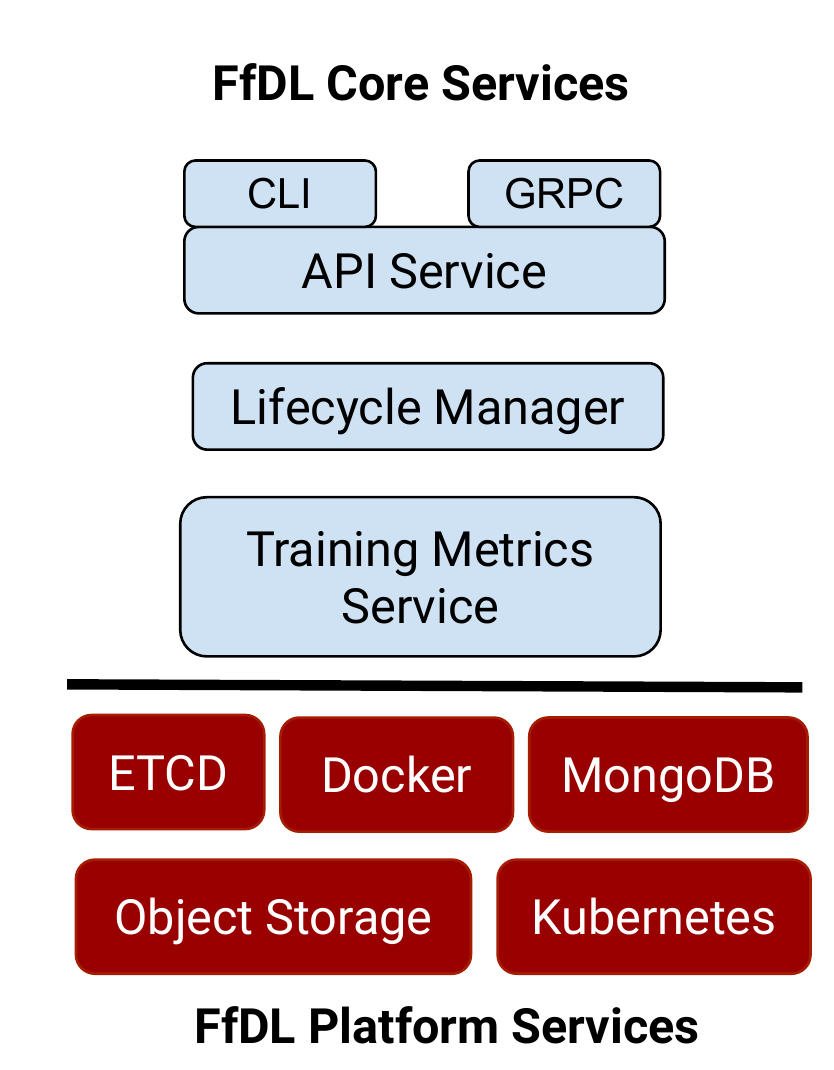}
    \caption{\DLaaS architecture illustrating its core/major components.}
    \label{fig:dlaasarch}
\end{figure}

\begin{figure}[htb]
    \centering
    \includegraphics[width=0.8\columnwidth]{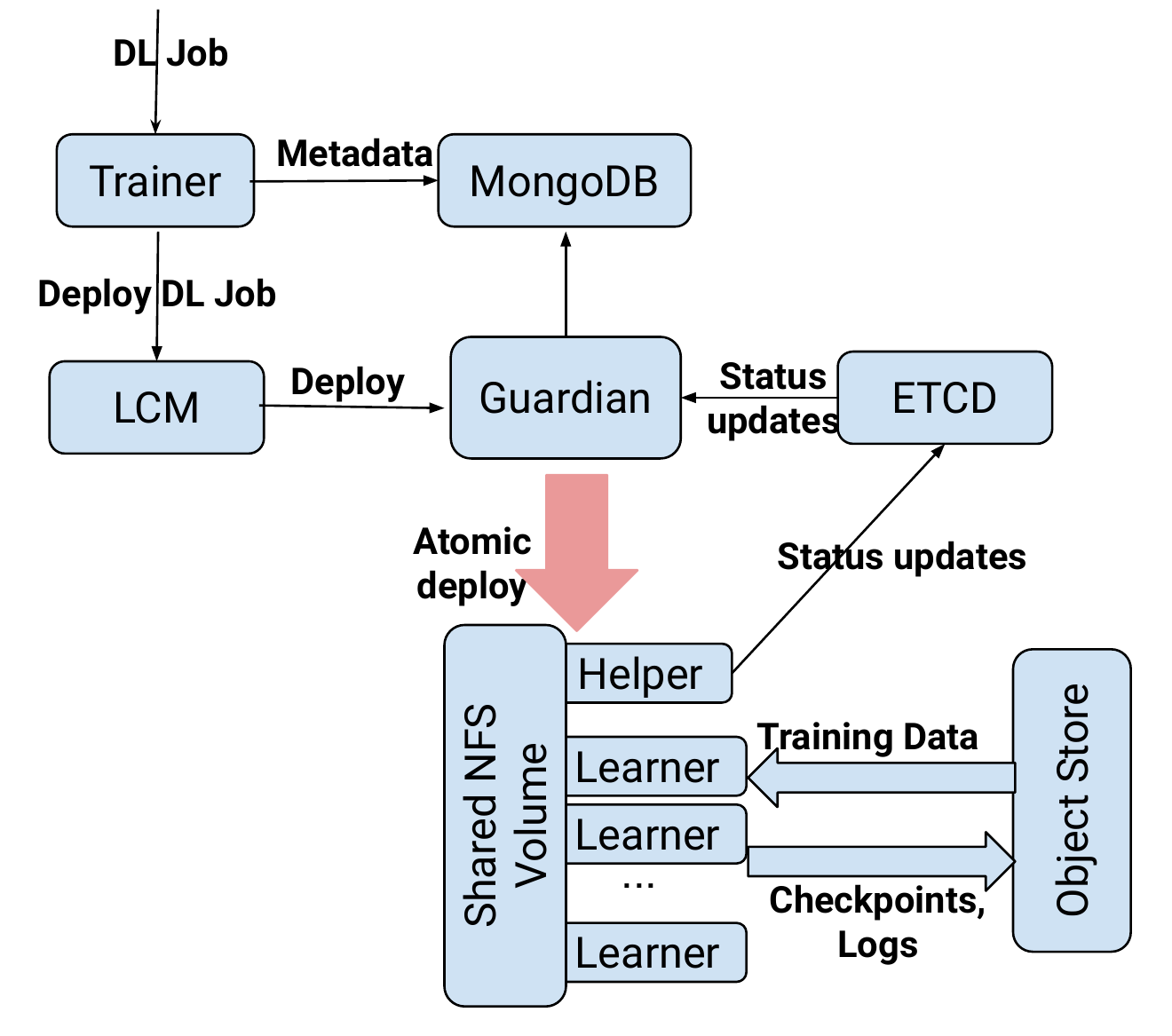}
    \caption{\DLaaS job deployment, execution and monitoring.}
    \label{fig:dlaasjobflow}
\end{figure}

\subsection{Flexibility through Containerization}

\DLaaS \emph{has to} support several popular DL frameworks such as TensorFlow, PyTorch, Torch, Caffe, and Horovod. 
To this end, \DLaaS uses container technologies 
(Docker~\cite{docker}) to abstract away framework specific details, and simplify the management of DL jobs.

Specifically, \DLaaS takes user code, and automatically instantiates Docker containers 
(from its library of Docker images) containing all necessary packages to run the training job.  
A DL training job typically consists of a set of learning processes (``learners''), 
with each learner running in a Docker container using one or more GPUs.  Each learner container 
consists of a framework Docker image (Caffe, TensorFlow, etc.) instantiated with user code and 
credentials (certificates, private keys, etc.) used for communication, downloading data, and 
storing results.  A distributed job may also include one or more parameter servers if the framework/user 
includes them; parameter servers are also containerized.

Communication and synchronization methods internal to frameworks are preserved while the job 
is executed in \DLaaS and are part of the framework Docker image.  This means that \DLaaS\ does 
not impose any requirements on the communication protocols, such as MPI or gRPC, used within a job.
\shep{FfDL simply requires data scientists to provide their existing code, 
command to execute said code, location of data, credentials to access said data and 
store results, number of learners, and the resources (GPU, CPU \& RAM) needed per learner. These items are 
described in a manifest file or as part of an API call; so the abstraction is close to 
a ``natural language'' DL job description. Data scientists simply get their training logs and results back;
and do not have to understand the internals of FfDL}.

\subsection{\DLaaS\ Architecture and Components}

\DLaaS is architected as a set of loosely coupled microservices, written in Go, and communicating 
over gRPC APIs, as shown in Figure~\ref{fig:dlaasarch}. Each microservice can be independently 
maintained, updated, deployed and scaled.  \DLaaS\ is composed of the following core 
microservices: (1) the API server (2) the Lifecycle Manager and (3) the Training Metrics Service.
\DLaaS has five dependencies: Kubernetes~\cite{kubernetes}, MongoDB~\cite{mongodb}, Object Storage Service, ElasticSearch, and 
etcd~\cite{etcd}.

\DLaaS employs Kubernetes (K8S) for container orchestration and cluster management. 
Each \DLaaS microservice is also (Docker) containerized and executed using a K8S abstraction 
called \emph{replica set}, with their APIs exposed using the K8S \emph{service} abstraction. 
Using replica sets enables us to replicate each microservice, and scale the number of replicas 
in response to cluster size and changes in workload. MongoDB is used to store job 
metadata (identifiers, resource requirements, user ids, etc.), as well as job history (for a specific user or client 
organization) and other business artifacts. Data stored in MongoDB is long lived (few months), and spans several DL jobs. 
The \DLaaS microservices 
coordinate with executing jobs using etcd, a fast and highly available key-value store.
Data stored in etcd is small (< 1KB) consisting of status updates and messages; and is short lived
(a DL job's data is erased after it terminates). We preferred to use etcd over MongoDB for coordination
because it is much faster and has some abstractions that MongoDB lacks, like leases on keys and 
fine grained support for ``streaming watches'' at the level of a single key. 
Both MongoDB and etcd are also replicated for high availability.


The \DLaaS API microservice handles all the incoming API requests, including load balancing, 
metering, and access management.  It exposes both a REST and gRPC API endpoint. The API 
service instances are dynamically registered
into a K8S service registry that provides load balancing and fail-over support 
for incoming API requests. 
When a job deployment request arrives, the API layer
stores all the metadata in MongoDB before acknowledging the request. 
This ensures that submitted jobs are never lost in case failures happen during 
job deployment, or a catastrophic failure temporarily takes down all machines in the 
cluster and all of \DLaaS\ core microservices. The information about a job in MongoDB is
continuously updated throughout its execution, including the current job status and job status history.
In the case of a failure that necessitates that the entire job be restarted (as opposed to a single component of a job), information stored in MongoDB can
be used readily without the need for user intervention.

The Training Metrics Service is responsible for collecting metrics about both the training jobs and \DLaaS microservices. This includes things like memory and network usage, number of times microservices fail and recover, and frequency of connectivity issues.  It also helps in streaming training logs from jobs to be indexed and stored in ElasticSearch/Kibana for easy debugging.

\subsection{Reliable Job Deployment}

The API layer submits every job to the Lifecycle Manager (LCM) microservice.  
The LCM is responsible for the job from submission to completion or failure. This includes 
deployment, monitoring, garbage collection, and user-initiated termination of the job.
The LCM performs some of these functions directly, but creates \emph{delegates} for
others, both for scalability as well as to avoid becoming a single point of failure.

Deploying a DL job is a multi-step process, involving placement on an appropriate 
cluster node with available GPUs, setting up network (MPI) interconnections, and provisioning 
shared volumes and credentials to access data. Users and cluster administrators
 require that provisioning of
DL jobs be atomic, either the whole job is provisioned with the requisite 
resources or none. In the case of an LCM replica crash in the middle of provisioning,
there should not be an inactive job component with allocated resources (i.e. a \emph{zombie}).
Consequently, if the LCM directly performs this multi-step job provisioning process, it has 
to store a lot of state, which goes against the microservices design philosophy. 

To mitigate this, the LCM launches a delegate for atomic deployment and further monitoring
of each DL job. If the delegate crashes for 
any reason (such as an OS, Docker, K8S, or machine failure),
K8S will automatically restart it and execute it again.
The LCM simply instantiates this delegate called the \emph{Guardian} 
with all the metadata of the DL job. The Guardian is a \DLaaS component created on 
the fly as a K8S Job for every DL job. Its creation is a very quick 
(less than 3s in our experiments) single step process. The Guardian then executes 
the multi-step process of actually deploying the DL job by further interacting with K8S. 
Some of the key steps in this workflow are instantiating Docker containers 
(corresponding to the DL framework used such as TensorFlow or PyTorch) with training 
parameters and user code, setting up shared NFS volumes to monitor training progress, 
and applying K8S policies to restrict network access from the learner in a multi-tenant environment. 
If the Guardian crashes in the middle
of a job deployment, K8S is guaranteed to restart it. The restarted Guardian will \emph{roll back}
the previous partially deployed DL job and start a fresh deployment process. In the presence of
persistent failures, this process will be repeated for a (configurable) number of times
before the Guardian gives up and marks the DL job in MongoDB as FAILED. Once a DL job is successfully deployed, the Guardian is then responsible for
monitoring its progress.

\subsection{Job Scheduling -- \textsc{Spread} vs. \textsc{Pack}}
\label{sec:scheduling-spreadvspack}
Our first prototype of \DLaaS used Kubernetes default scheduler as is. However, the very first
set of tests that we performed on a development cluster of 20 K80 GPUs raised several
issues. First, the default placement policy of Kubernetes pods was \textsc{Spread} -- due to the 
fact that Kubernetes was optimized to run workloads like websites, databases, etc. \textsc{Spread}
distributes pods over the cluster, and avoids placing two pods which are replicas of the same 
workload on the same physical machine. However, in the case of DL training jobs, \textsc{Spread}
caused two problems: (i) it increased communication costs due to learners being spread throughout 
the cluster, and more importantly, (ii) it increased \emph{fragmentation}. To see why \textsc{Spread}
increases cluster fragmentation, consider the following simple example: 4 DL training jobs, each with
1 learner and 1 GPU/learner arrive at a cluster with 4 machines and 4 GPUs/machine. \textsc{Spread}
would distribute the jobs over all 4 machines. Now, we have 4 machines, each with 3 GPUs free. 
If a new job arrives with 1 learner and 4 GPUs/learner, it cannot be scheduled due to lack of
resources. 

So, we made a decision to use the \textsc{Pack} placement policy, where pods from
a DL job are packed (``crammed'') into as few physical machines as possible.
We implemeted an extension to the K8S scheduler to support \textsc{Pack}.
In the scenario outlined above, \textsc{Pack} would place all four jobs on the same 
machine, leaving three machines free with 4 GPUs/machine. The cluster would not 
only be able to support the new job with 4 GPUs/learner, it would be able to support
two more such jobs. Section~\ref{sec:spreadvspack} presents an empirical
comparison of \textsc{Spread} vs. \textsc{Pack} in the context of cluster fragmentation

\textsc{Spread} is useful for fault tolerance of traditional workloads, where 
spreading replicas throughout the cluster helps survive catastrophic failures.
In the case of DL jobs and heavily loaded clusters, however, \textsc{Spread} may result in the failure of
multiple jobs when a machine crashes; each of the failed jobs either needs to
restart and recover from a checkpoint or wait for the failed learner to recover and
rejoin. With heavily loaded clusters, \textsc{Pack} may impact fewer jobs. 

\subsection{Job Scheduling -- Gang Scheduling}\label{sec:gangscheduler}

When we started testing our implementation of \DLaaS\ on moderately sized clusters (40 GPUs),
we observed that under heavy load, in the presence of distributed DL jobs,
 a nontrivial percentage of learners became \emph{temporarily deadlocked}.
Upon investigation, we determined that this is a shortcoming of the K8S
scheduler -- which scheduled individual pods, rather than scheduling
all pods that belong to a DL job holistically, as a group/gang.
To see why this can be problem, consider a cluster with 4 machines,
and 2 GPUs/machine. Assume 4 synchronous DL jobs with 2 learners/job and 2 GPUs/learner
arrive. Ideally, the K8S scheduler should schedule two jobs and queue the remaining two.
But, the K8S scheduler doesn't consider one whole job while scheduling, it considers
each of the \emph{learner pods} individually. Consequently, one learner pod from each of the
four jobs might be scheduled on each of the four machines, and the other four pods queued. None of the learners
can make progress because they are waiting for the other learner.
They are \emph{temporarily deadlocked} and hoarding GPUs until at least one of them gives up and fails. 
We demonstrate this empirically in Section~\ref{sec:evalgangscheduling}.

This can be avoided if all components of a DL job are scheduled as a \emph{gang}.
In general, a DL job comprises a collection of Kubernetes sets (e.g. stateful sets), 
where each set is a collection of homogeneous pods where tasks, such as learners and parameter servers, run.
In addition to pods, the DL job deployment includes other entities such as volumes and networking interfaces.
We will refer to a scheduler whose function is to place all pods that belong to a 
DL job onto nodes in the cluster holistically, as a gang scheduler.
The gang scheduler matches the requested resource demands of all the pods in a DL job 
(e.g. CPU, memory, GPU, and storage) with the available resources on the nodes, finding a set of nodes
on to which pods are placed.
This placement problem is known as the (multi-dimensional) bin packing problem, which is NP-hard.
This placement problem is restated as an assignment optimization problem as follows --
Given a collection of {\it logical entities} (LE), corresponding to pods, and a collection of {\it physical entities} (PE), corresponding to nodes, with a set of constraints related to resources and topologies, such as pairwise-constraints, as well as an objective function, such as load balancing, find an optimal solution, where each LE is assigned to a PE.

Our gang scheduler uses the Biased Sampling Algorithm (BSA)~\cite{Tantawi15,Tantawi16}, which we believe
is a promising heuristic approach to solve this assignment optimization problem.
A complete description of BSA is beyond the scope of this paper, and is available in \cite{Tantawi15,Tantawi16}.
\shep{We provide a high-level description of how BSA was adapted to work as a K8S gang scheduler.
When a pod is not bound to a node, the K8S default scheduler assigns a node to the pod by (1) filtering the nodes that satisfy the pod resource requirements and other predicate constraints, (2) ranking the candidate nodes based on priority functions, and (3) selecting the node with the highest rank.
We modified this process by first identifying the gang information, namely gang name and gang size, of the pod.
Such information is readily available from the pod owner, such as {\it replicaSet} or {\it statefulSet}.
Without loss of generality, let us assume that pods in the group are homogeneous.
Then, the {\it logical entities} are the all pods in the gang and the {\it physical entities} are the nodes in the cluster, given its current state.
Since in a DL platform, GPU is typically a scarce resource, the objective is to pack GPU resources.
The default filtering and ranking steps are mapped onto node preferences (or biases) for placement of the pods.
Then, an instance of a placement problem is created and given to the BSA algorithm library to solve.
At the scale of typical DL clusters, the solution space is combinatorially explosive, so BSA uses importance sampling techniques to probabilistically sample nodes from the cluster. BSA reuses cluster and node information stored by K8S.
BSA biases the sampling towards nodes that both satisfy the constraints and optimize the objective function. 
The returned vector of nodes is then used to bind nodes to pods.
In case pods in the group are not seen by the scheduler yet, (corner case, when the pods are still being instantiated by K8S) 
the node assignment is stored as a reservation for future pod arrivals.
}

Though the gang scheduler required effort to implement, it guaranteed zero temporary deadlocks irrespective of the workload as demonstrated in Section~\ref{sec:evalgangscheduling}.

\subsection{Job Scheduling -- Putting it all together}

\shep{Baseline job dispatching in FfDL is first come first served (FCFS). GPUs currently do not have 
support for fine grained sharing; time sharing is supported (not with high efficiency), but space sharing between different workloads is currently not supported. Hence, we do not overcommit GPUs while scheduling DL workloads. FCFS with gang scheduling and PACK works quite well given no overcommitment. The corner case when multiple jobs arrive at the same instant, the FCFS conflict is resolved by picking the largest gang (job) first. To summarize, jobs are considered in FCFS, each job is scheduled as a gang (or queued if not possible), and the gang is PACKed. }

\shep{Given that there is no overcommitment, admission control (AC) becomes necessary; there is a component above FfDL that performs AC -- based on quotas for internal users, and based on pricing/agreements for external users. 
Fair sharing doesn't work well -- some users/customers seem to be/become more important than others. 
Currently, the AC component also pre-empts 2 job types as necessary: 
(1) free users during heavy load, and (2) user A exceeded their quota; their job was scheduled because user B wasn't using their quotas; user B subsequently wants to use his quota. More advanced priority management (PM) based on demand-driven pricing for external users, and exponentially decreasing priorities for heavy internal users are part of ongoing work. AC and PM policies are typically different for internal, on-premise and public cloud users and are sometimes decided by business priorities; hence they are logically external to FfDL.}

\subsection{Scalability and Elasticity}

The microservice-based design is the primary contributor to the scalability of \DLaaS.  Each of the microservices is specialized to perform one core function at any given time. Each microservice is replicated, with the number of replicas chosen based on the size of the cluster as well as the expected number of concurrent training jobs. The replicated microservices are K8S deployments or replica sets, which means that gRPC requests to them are automatically load balanced by K8S among the available replicas. The replica set for each microservice can also be elastically scaled as physical machines are added to or removed from the cluster. The use of the Guardian for job deployment ensures that \DLaaS can handle thousands of deployment requests concurrently. Guardians consume only a fraction of a CPU and need little RAM. 

\DLaaS also monitors the usage of the cluster in terms of the percentage of GPUs currently
allotted to jobs. This is done both by querying K8S as well as from the job metadata
(including resource requests) stored in MongoDB. Given that context switching severely 
affects performance while training neural networks, we do not do time division multiplexing
of GPUs while scheduling training jobs.
Depending on usage, a systems administrator or an automated process can 
scale the size of the cluster either to accommodate more jobs or to limit wastage of resources.

\paragraph{Mounted object store} Training data and models can be large, in the order of gigabytes or terabytes, and often stored in a storage backend such as a Object Storage. \DLaaS can mount remote data in the learner container, so DL frameworks can access training data as though it were on the local filesystem. A driver streams files on demand and caches them so they can be reused across training epochs and jobs. This is an important optimization for several use cases, including those detailed in this paper.

\subsection{Robustness}

\paragraph{Detecting Failure or Completion of Learner Processes.} The Gu\-ardian uses K8S \emph{stateful sets} to deploy a DL Job. This enables \DLaaS to create replicated learners (for distributed training) and is well suited for DL frameworks like Horovod and distributed TensorFlow. For each DL job, the Guardian also creates a separate \emph{helper K8S pod} using the K8S \emph{Deployment} abstraction, which contains a number of ``helper'' containers: \emph{load-data} and \emph{store-results} to load and store data, \emph{log-collector} to process logs, and \emph{controller} to orchestrate the job. The helper pod remains isolated from the learner pods, but both share a common NFS filesystem, mounted by the Guardian using a K8S persistent volume claim.  The shared NFS volume enables the controller container running separately in the helper pod to monitor the execution and exit status of the learner processes and detect both their completion and failure by reading their output and process exit statuses redirected to a file. 

\paragraph{Reliable Status Updates.} In addition to detecting learner completion and failure, the controller can read the status/output of the load-data and store-results containers because all the helper and learner containers share a common file system. To reduce coupling between \DLaaS components and ensure reliable status updates, we employ the etcd key-value store to co-ordinate between the controller and LCM/Guardian.  etcd itself is replicated (3-way), and uses the Raft consensus protocol to ensure consistency.  The controller records the current status of each learner in etcd, where it is read by the Guardian.  The Guardian aggregates the statuses of each learner to record the overall status of the job in MongoDB, from where the user can read it through a REST/gRPC API call to \DLaaS. Using etcd makes status updates resilient to crashes of both the controller/helper pod and crashes of the Guardian. Using NFS makes status updates resilient to controller crashes; K8S will restart the controller which can read current status and previous statuses from NFS.

\paragraph{Node/Container Crashes.} Situations where the learner fails, i.e., by writing an appropriate exit code to NFS, can be detected by the controller. However, DL job \emph{crashes} due to node/container crashes are handled by K8S. Crashed learners will be restarted automatically by K8S, because learners are deployed as stateful sets. A recovered learner can continue training either (1) from the latest checkpoint, or (2) in the case of distributed DL jobs, by rejoining other learners and getting the latest neural net parameters from a parameter server (if the DL framework supports this). The work lost due to a crash is determined by the checkpointing interval.

\paragraph{Checkpointing} Given the long running nature of DL training jobs, checkpointing is vital. For jobs running in FfDL, checkpoints are
stored in a mounted object store. The checkpointing interval depends on the tolerance level of the user to failures, i.e., how many hours of work the user is willing to lose in the event of a failure. All user-driven checkpointing features of frameworks like Caffe, TensorFlow and PyTorch, including periodic checkpoints, are supported by FfDL. DL frameworks like Caffe support ``higher-level'' checkpointing (like defining checkpoint
intervals in a manifest file). For such frameworks, in the event of a failure, after the training pod is restarted, 
a FfDL component running inside the pod searches the object store bucket for
the latest checkpoint and uses that to resume training (i.e., populates the path of the latest checkpoint in the command to resume training). 
For other frameworks (like TensorFlow and PyTorch), where creation and loading of checkpoints is part of user-code, upon recovery from
failure, FfDL automatically mounts the object store bucket containing the checkpoints to the recovered pod. In this case, actual loading of the checkpoint and its use is the responsibility of user code. Checkpointing also enables FfDL to support user-driven HALT/RESUME for hyperparameter tuning.

\paragraph{Security and Isolation}
All communication between \DLaaS microservices is encrypted. Data transfers to and from the remote data store can be optionally encrypted per user needs. Containerization using
Docker ensures limited isolation; further isolation is provided through Kubernetes network
policies which are defined by the Guardian for every deployed job. These policies ensure that all jobs are isolated from each other.

\section{Lessons learned}
\label{sec:lessons}

We have been involved in every stage of building \DLaaS, including gathering requirements, developing the code, testing with sponsor users, operating in production, and supporting users. There are number of lessons we have learned during this journey.

One recurring observation was that much of the deep learning stack is not designed for the cloud. Despite GPUs inherently supporting
parallelized workloads, they cannot be \emph{space-shared}, i.e., if a GPU has 2000 cores, two applications, each needing
1000 cores cannot securely share the GPU simulataneously. This severely limits resource optimization; we have observed
that although DL jobs are GPU-hungry, they often do not utilize 100\% of the GPU. Although GPUs can be virtualized, this only enables
them to be \emph{time-shared} between applications, which is not very useful for DL workloads due to the costs associated with
context switching. Plus, it was necessary to securely wipe GPU memory between jobs. Furthermore, cluster managers do not have facilities to detect GPU failures and migrate workloads, which became a problem when we noticed that faulty GPUs were not uncommon. The deep learning frameworks were also mostly designed and optimized for HPC environments. There was little to no support for fault-tolerance, elasticity, resource heterogeneity, resource sharing, and security. Much of the code in \DLaaS is to support these requirements at the middleware layer in order to avoid having to modify the DL frameworks.

Impedance mismatch between DL and cloud goes both ways: the cloud software stack was not designed for DL workloads. For example, Kubernetes is typically used for Web applications and microservices that are built to scale horizontally. DL jobs, on the other hand, are long running and cannot natively be migrated, scaled out, or suspended. In particular, many distributed DL jobs require all workers to be running to make progress; a partially deployed set of workers will render the job idle and waste resources. The schedulers in cluster managers do not account for this, and we had to develop custom Kubernetes gang scheduling logic as described in Section~\ref{sec:gangscheduler}.

We were caught off guard with some changing requirements. The early sponsor users of the system were service operators who needed a highly available and scalable service to train DL models. This results in a high volume, automated, and relatively homogeneous set of jobs. The service was then increasingly used by data scientists who wanted as much control over their \DLaaS jobs as with their local machine. This included the ability for jobs to download datasets or Python packages from the public Internet, the need to connect to running jobs in order to interactively debug models, and requests to stream logs to local scripts in order to monitor the progress of jobs. While we were able to modify \DLaaS to accommodate these different usage patterns, some early users were frustrated by the initial user experience.

\DLaaS was designed with multiple measures to protect against malicious and inadvertent attacks. This includes physical isolation of \DLaaS micoservices, container isolation for user code, network isolation with restrictive communication policies, two-way authentication between all microservices, encryption of the control and data planes in flight and at rest, aggressive purging of potentially sensitive information in logs and databases, and live cluster and job monitoring. This defensive security posture paid dividends on multiple occasions. Once, we inadvertently leaked some of the certificates used by the \DLaaS microservices outside the core development team, but were confident enough in the other defenses to leave the service running while we rolled out new certificates. Also, when faced with the interactive user requirements above, the defense-in-depth approach made it easier to justify relaxing some of the defensive measures to improve usability.

Some of the initial storage design decisions turned out to be expedient but not desirable. For example, the use of dynamically provisioned NFS volumes per job for training data was quick to implement and met the security requirements, but did not scale. First, provisioning NFS volumes was slow and often failed under high load. Attempts to address this with a microservice to pre-allocate and manage a pool of NFS volumes only increased the complexity of the system. We settled on a simpler and more robust solution of mounting object storage buckets, hiding the streaming of data from a remote cloud service under a file system interface. We continued to use NFS volumes as a secure communication channel between untrusted containers, but the dependency on NFS could have been removed entirely with a bit more redesign of this aspect. We should have also spent more effort earlier on optimizing the storage layer. DL training sets can be very large, but the same datasets are often used across jobs, and an intelligent caching layer tuned to DL access patterns could have significant cost and performance improvements.

We expected there to be rapid change in types of GPUs and the DL frameworks that \DLaaS would need to support, and designed the core system to be agnostic to both GPUs and frameworks. Despite this not being an initial requirement from sponsor users, it turned out to be a good decision since we have had to support multiple generations of GPUs and almost weekly updates to the DL frameworks. Being DL framework agnostic increased the complexity of the middleware but in retrospect was absolutely necessary.

Supporting the logging generated by distributed DL jobs turned out to be surprisingly challenging. Initial designs had a microservice that would stream stdout logs from the container where a job was running, via a public API. This streaming traffic increased the load on \DLaaS, required complex logic to avoid gaps in logs under failure conditions, and increased the vulnerability exposure of the service by passing sensitive logs over yet another channel. Attempts to stream the logs to the user's object storage bucket were impeded by the lack of append operations in these services. Also many users wanted the ability to pipe logs to custom scripts, as they would for locally running jobs, and the \DLaaS logging APIs added a layer of complexity to their workflow. We developed solutions to make it easy to run custom parsers and scripts along with the jobs, but quickly realized that it would not be feasible to keep up with wide variety of specialized user requirements.

\section{Evaluation} \label{sec:evaluation}

\begin{table*}[htb]
\small
    \begin{tabular}{|c|c|c|c|c|c|c|c|c|}
    \hline
    \multicolumn{9}{|c|}{VGG-16~\cite{vgg16} Image Processing Benchmark on Caffe v1.0} \\
    \hline
    Config. & 1L $\times$ 1GPU/L &  1L $\times$ 2GPU/L &  1L $\times$ 4GPU/L & 2L $\times$ 1GPU/L &  2L $\times$ 2GPU/L &  2L $\times$ 4GPU/L &  4L $\times$ 2GPU/L & 4L $\times$ 4GPU/L \\
    \hline
    Decr. in Perf. & 3.29\% & 0.34 \% & 5.2 \% & 3.76\% & 2.45\% & 4.76\% & 3.2\% & 5.35\% \\
    \hline
    \multicolumn{9}{}{}
    \end{tabular}
    \begin{tabular}{|c|c|c|c|c|c|c|c|c|}
    \hline
    \multicolumn{9}{|c|}{Inception V3~\cite{inceptionv3} Image Processing Benchmark on TensorFlow v1.5} \\
    \hline
    Config. & 1L $\times$ 1GPU/L &  1L $\times$ 2GPU/L &  1L $\times$ 4GPU/L & 2L $\times$ 1GPU/L &  2L $\times$ 2GPU/L &  2L $\times$ 4GPU/L &  4L $\times$ 2GPU/L & 4L $\times$ 4GPU/L \\
    \hline
    Decr. in Perf. & 0.32\% & 4.86 \% & 5.15 \% & 1.54\% & 3.65\% & 3.96\% & 4.2\% & 4.97\% \\
    \hline
    \end{tabular}
    \caption{Performance overhead of \DLaaS vs. Bare Metal servers on popular Image Processing Benchmarks. Performance is quantified as images processed/sec for training. Caffe v1.0 and TensorFlow v1.5 were used.}\label{fig:overhead-sl}

\end{table*}

\begin{table}[htb]
    \centering
    \begin{tabular*}{0.48\textwidth}{c|c|c|c|c}
    \hline
    Benchmark   & Frame-  &  \# PCIe  & GPU & Difference in  \\
                &      work      &    GPUs  & Type & Performance \\
    \hline
    Inceptionv3~\cite{inceptionv3} & TF & 1 & P100 & 3.30\% \\
    \hline
    Resnet-50~\cite{resnet50} & TF & 1 & P100 & 7.07\%  \\
        \hline
    VGG-16~\cite{vgg16} & TF & 1 & P100 & 7.84\% \\
    \hline
    InceptionV3~\cite{inceptionv3} & TF & 2 & P100 & 10.06\% \\
    \hline
    Resnet-50~\cite{resnet50} & TF & 2 & P100 & 10.53\% \\
        \hline
    VGG-16~\cite{vgg16} & TF & 2 & P100 & 13.69\% \\
    \hline
    \end{tabular*}
    \caption{Performance overhead of \DLaaS vs. NVIDIA DGX-1 bare metal server on TensorFlow HPM benchmarks~\cite{tensorflowcnn}.  Performance is quantified as images processed/sec for training. }
    \label{fig:overhead-dgx}
\end{table}

\begin{table}[htb]
    \centering
    \begin{tabular}{c|c}
    Component & Time to recover \\
              & from crash failure \\
              
    \hline
    API  & 3-5s \\
    LCM  & 4-6s \\
    Guardian & 1-2s \\
    Helper & 3-4s \\
    Learner & 10-20s \\
    \end{tabular}
    \caption{Time taken to recover from crash failures, by component.}
    \label{fig:recoverytime}
\end{table}

\begin{table}[htb]
    \centering
    \begin{tabular}{|c|c|c|}
    \hline
    CPU-threads & thpt-1P100 &  thpt-1V100  \\ \hline
      2     & 65.96 &  106.46  \\
        4    & 66.14 &  106.5  \\
        8   &  65.67 &  107.24 \\
        16  &         & 107.45  \\
        28  &         &  107.47  \\
        \hline 
    \end{tabular}
    \caption{ Throughput (images/sec) scaling of VGG-16/Caffe with CPU threads for learners with 1 P100 and 1 V100 and batch size 75.}
    \label{tab:caffe}
\end{table}

\begin{table}[htb]
    \centering
    \begin{tabular}{|c|c|c|}
    \hline
        GPU-type & CPU & memory (GB) \\
        \hline   
        1-K80 & 4 & 24   \\
        2-K80 & 8 & 48 \\
        4-K80 & 16 & 96 \\
        1-P100 & 8 & 24 \\
        2-P100 & 16 & 48 \\
        1-V100 & 26 & 24 \\
        2-V100 & 42 & 48 \\
        \hline 
    \end{tabular}
    \caption{ T-shirt size recommendation for \DLaaS jobs.}
    \label{tab:tshirt}
\end{table}

\begin{table}[htb]
    \centering
    \begin{tabular}{|c|c|c|c|}
    \hline
    CPU-threads & InceptionV3 & Resnet-50 & VGG-16 \\ \hline 
        16 & 217.8 (86.8\%) & 345.3 (93.3\%) & 216.2 (98.7\%)\\
        28 & 223.6 (90.5\%) & 345.8 (92.7\%) & 216.2 (97.3\%) \\
        \hline 
    \end{tabular}
    \caption{ Throughput (images/sec) scaling of TensorFlow with CPU threads for learners with 1 V100 and batch size 128. Also shown in parentheses is the GPU utilization.}
    \label{tab:tf}
\end{table}

The migration to a cloud based DL platform from bare metal servers is 
influenced by the cost-benefit tradeoffs. In particular the benefits of having a 
dependable, fault-tolerant, and cost-effective execution environment can overshadow 
a slight degradation in performance.  In this section, we demonstrate empirically 
that the dependability features of \DLaaS and execution in a containerized environment have minimal impact on performance
of DL training jobs. We illustrate this by using several DL benchmarks~\cite{dlbenchmarks}
(VGG-16~\cite{vgg16}, Resnet-50~\cite{resnet50} and InceptionV3~\cite{inceptionv3}),
three different PCIe-based GPU types (K80~\cite{k80} and P100~\cite{p100} and V100), and 
two different DL frameworks (Caffe v1.0~\cite{caffe} and TensorFlow v1.5~\cite{tensorflow}).

\subsection{Runtime overhead}

For our first set of measurements, we compare \DLaaS with directly executing the 
benchmarks (non containerized) on bare metal machines manually on \IBM datacenters. 
In both cases a 1GbE interconnect was used, and training data was stored in an  
Object Storage Service (OSS) hosted on \IBM datacenters. Results are illustrated in Table~\ref{fig:overhead-sl}.
In the overhead measurements of Table~\ref{fig:overhead-sl}, we vary both the DL job configuration
as well as the framework used. We consider both single learner as well as distributed DL 
jobs. We vary the number of learners from 1 to 4, and the number of GPUs/learner from 1 to 4 as well.
From Table~\ref{fig:overhead-sl}, we observe that performance overhead induced by \DLaaS is minimal (up to $\approx$ 5\%).
\shep{We believe the source of the this overhead is predominantly (1) Docker (very low but non-zero) (2) network virtualization and network security policies and (3) a driver to mount Cloud Object Storage buckets (where the training data resides) onto Kubernetes pods.}

For the second set of measurements, we compare \DLaaS to NVIDIA's specialized hardware, DGX-1~\cite{dgx}, which incurs $\approx$2-3$\times$ additional costs compared to off-the-shelf hardware (such as from \IBM~\cite{ibmcloudgpu}). 
DGX-1 has advanced hardware (NVLink and High Bandwidth Memory), and is expected
to have higher performance than \DLaaS. However, we observe from Table~\ref{fig:overhead-dgx}, that degradation in performance, though nontrivial is only modest (up to $\approx$ 15\%).

Finally, our cloud-native design and implementation has ensured that \DLaaS remains loosely coupled and each component can fail independently of the other. Within a DL training job, a learner can crash and be restarted by K8S independently of the helper. Guardians can crash/restart independently of the LCM and API, and so on. Time taken for each component to restart is minimal and illustrated in Table~\ref{fig:recoverytime}. These times were calculated by manually crashing various components (using the K8S \texttt{kubectl} tool) and measuring the time taken for the component to restart. Learners take longest to restart because binding to the Object Storage Service and persistent NFS volumes takes longer, and
FfDL microservices take the shortest time because they are stateless.

\subsection{Comparing \textsc{Spread} vs. \textsc{Pack}}~\label{sec:spreadvspack}

\begin{figure}[htb]
\subfigure[Arrival of jobs by day over a 60 day period at a production cluster~\label{fig:spreadpack-jobarrival}]{
\includegraphics[width=0.9\columnwidth, keepaspectratio=true]{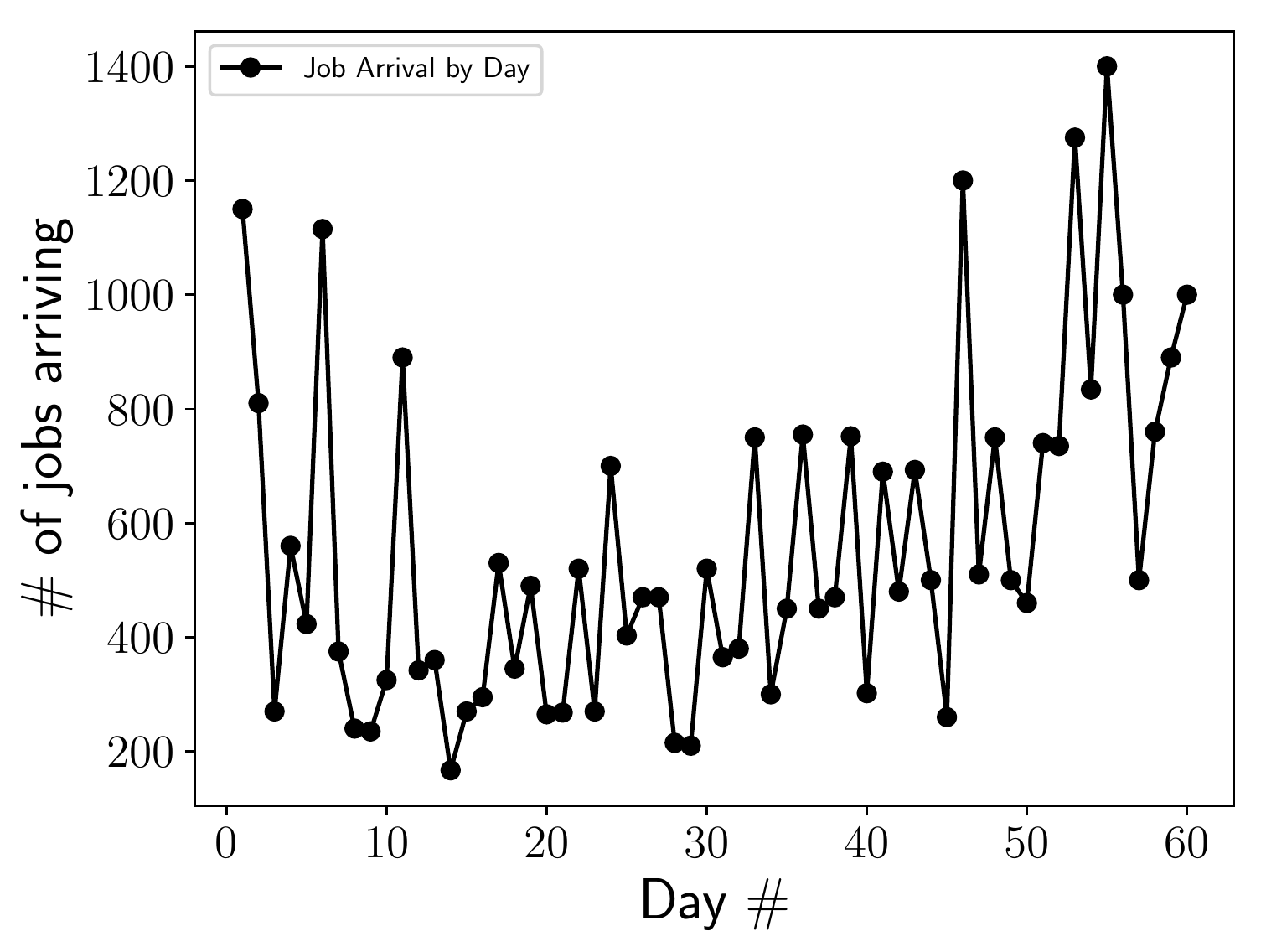}
}
\subfigure[Percentage of arriving jobs that would be queued for over 15 mins, in \textsc{Spread} vs. \textsc{Pack}~\label{fig:spreadpack-queuedjobs}]{
\includegraphics[width=0.9\columnwidth, keepaspectratio=true]{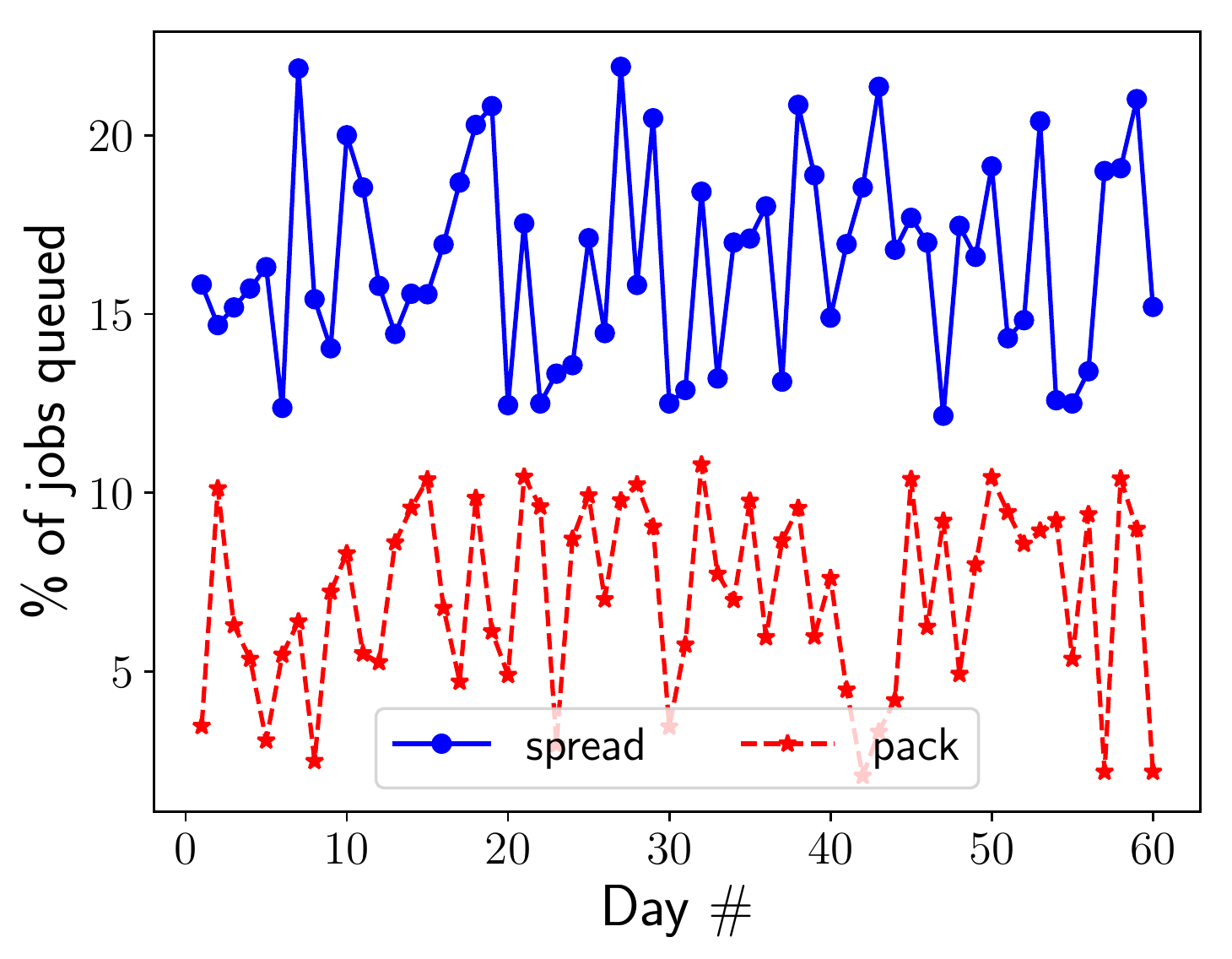}
}
    \caption{Spread vs. Pack scheduling performance and job arrival pattern}
\end{figure}

As described in Section~\ref{sec:scheduling-spreadvspack}, we discovered fragmentation 
when we tested our first prototype of \DLaaS on a 20 GPU cluster. So we did not 
deploy \textsc{Spread} based job scheduling in production. But, to test our hypothesis,
we collected job arrival traces on a production cluster with 400 GPUs (180 K80s and 220 V100s)
over a 60 day period. From these traces, for each job, we obtained the arrival time,
duration, GPU requirements, etc. We then simulated the effect of using both \textsc{Spread}
and \textsc{Pack} to schedule these jobs, and measured the number of jobs that are
queued for more than 15 minutes because the requisite GPU configuration is unavailable.
We aggregate the results by day and plot the same in Figure~\ref{fig:spreadpack-queuedjobs},
correlating it with the job arrival rate by day in Figure~\ref{fig:spreadpack-jobarrival}.
From Figures~\ref{fig:spreadpack-jobarrival} and \ref{fig:spreadpack-queuedjobs},
we observe that \textsc{Pack} results in significantly fewer jobs queued for more than 15 
minutes -- over 3$\times$ fewer queued jobs. We use the 15 minute threshold because it emerged as
the threshold beyond which user satisfaction with the DL platform drops significantly.

\subsection{Need for Gang Scheduling}~\label{sec:evalgangscheduling}

\begin{figure}[htb]
\subfigure[CDF of the probability of temporarily deadlocked learners with and without gang scheduling~\label{fig:deadlocked-learners}]{
\includegraphics[width=0.9\columnwidth, keepaspectratio=true]{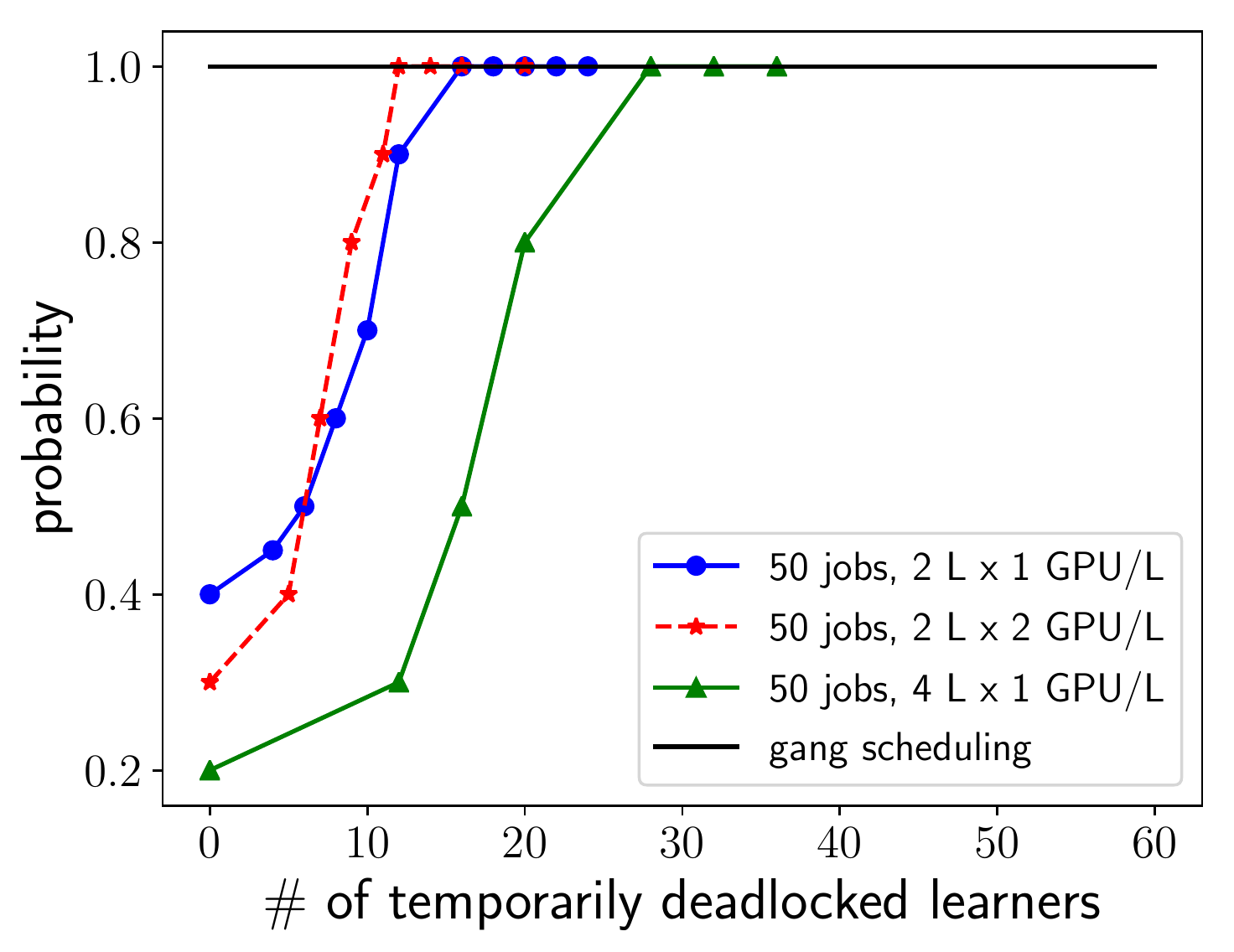}}
\subfigure[CDF of the probability of idle GPUs with and without gang scheduling~\label{fig:idlegpus}]{
 \includegraphics[width=0.9\columnwidth]{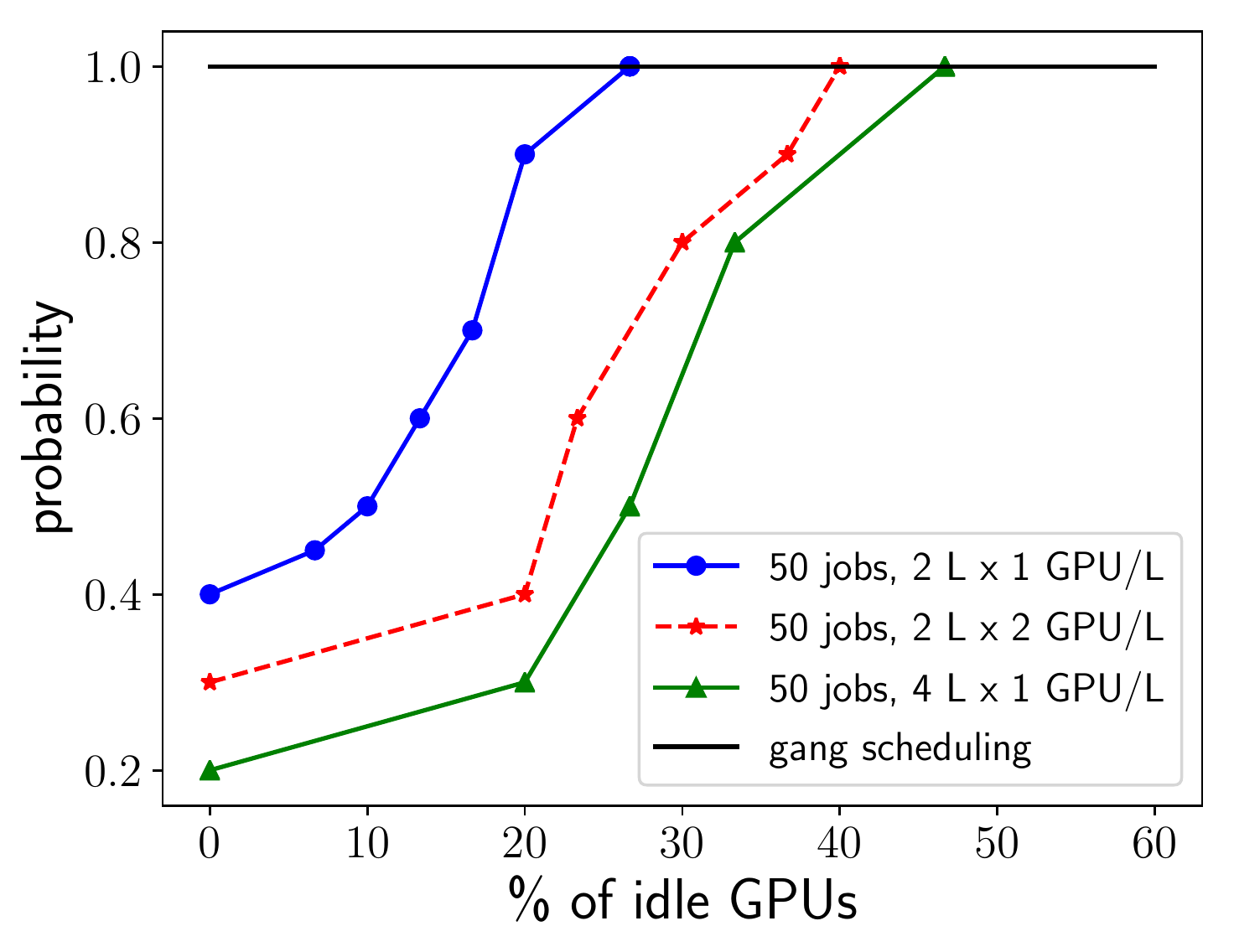}   
}
\caption{Need for gang scheduling}
\end{figure}

Next, we demonstrate the need for gang scheduling empirically using synthetic
workload with a cluster of 15 machines, with 4 K80 GPUs each (total of 60 K80 GPUs).
We deploy \DLaaS on this cluster with and without the gang scheduler.
We consider three workloads, of 50 synchronous DL training jobs each: (i) jobs with 2 learners, 1 GPU/learner,
(ii) jobs with 2 learners, 2 GPUs/learner and (iii) jobs with 4 learners, 1 GPU/learner.
These jobs are submitted concurrently to \DLaaS, which in turn, submits them to 
K8S for execution. We repeat each experiment 20 times, with and without gang scheduling,
 and measure the number of 
temporarily deadlocked learners in each run. Since the order in which learner pods are
queued by K8S for scheduling is non deterministic, the number of temporarily deadlocked
learners will not be the same for each run. Some runs may not have any deadlocks.
From the measurements over 20 runs, we compute the probability of deadlocks and 
idle GPUs, e.g., for the workload with 2 learners/job and 1 GPU/learner, if 2 runs out of 20 
resulted in 5 deadlocked learners, the probability of 5 deadlocked learners is $2/20~=~0.1$.
We plot the CDF of the probability of deadlocked learners and percentage of idle GPUs
in Figures~\ref{fig:deadlocked-learners} and ~\ref{fig:idlegpus} respectively.

Consider the first workload -- 50 jobs $\times$ 2 L/job $\times$ 1 GPU/L. This creates
a demand of 100 GPUs against a supply of 60 GPUs. Ideally, 30 out of the 50 jobs should get fully scheduled
for execution, and 20 jobs should be queued until the executing jobs finish. In other words,
a job should either be \emph{fully scheduled} or \emph{fully queued}. 
From Figure~\ref{fig:deadlocked-learners}, however, we observe that 
this happens only with a probability of 0.4 (40\% of the time). 
There are temporarily deadlocked learners 60\% of the time, though their  number
varies between 4 and 12. When a learner is temporarily deadlocked waiting
for other learners to be scheduled, the GPU held by the learner is idle
because training has not started. Figure~\ref{fig:idlegpus} plots the 
CDF of the probability of idle GPUs. We observe, for example, that 
for the first workload, GPUs are not idle only 40\% of the time, and the 
percentage of idle GPUs can reach as high as 46\%.

We also observe that the number of temporarily deadlocked learners and the percentage
of idle GPUs in the cluster increases as the jobs become more resource intensive and 
distributed. For the 4L $\times$ 1 GPU/L case, the percentage of idle GPUs reaches
46\% in some runs. This is clearly unacceptable, and was the reason why we invested
resources to develop a gang scheduler for K8S. The number of idle GPUs and the 
number of temporarily deadlocked jobs has been zero, for all runs with gang scheduling.

\subsection{Resource sizing for \DLaaS jobs}

Cloud based platforms should offer dependable performance, and users should get the expected performance based on their hardware choices. 
Since DL training jobs employ GPUs attached to a node, and are provisioned in a container, the amount of resources (memory and CPU) allocated to the container 
also affects the performance. Although the end users have flexibility to specify their own configuration for learner resources, \DLaaS provides guidelines to users 
on resource sizing for learner pods based on their GPU type. The goal is to dimension the CPU threads per learner to achieve close to 100\% utilization of the GPUs. 
and prevent performance degradation due to poor resource choices made by ``uninformed'' users. We selected the maximum batch size for each model and then increased the CPU threads to saturate the GPUs. While all models/frameworks benefit with increased CPU threads, the scaling was observed to be framework dependent. Table~\ref{tab:caffe} shows the scaling of performance with CPU threads for VGG-16/Caffe with P100 and V100 GPUs. Observe that Caffe models performance saturates at 4/8 CPU threads for
both P100 and V100 GPUs. 

Table~\ref{tab:tf} shows CPU scaling for TensorFlow models for V100 GPUs with batch size 128. Observe that TensorFlow models show benefits up to 28 CPU threads. We also observed that learner pod memory of around 9GB is sufficient for most of the jobs and this memory utilization does not depend on GPU type.  
Based on these findings we created ``t-shirt size'' of learners for \DLaaS jobs which are framework agnostic, for simplicity. Table~\ref{tab:tshirt} shows the learner t-shirt sizes for different
GPU types. These ensure that we are utilizing GPUs to the fullest and \DLaaS jobs performance is not bottlenecked due to CPU and memory of learners.

\shep{We note that T-shirt sizes were solely determined with the goal of saturating GPUs as described above.
In this sense it is conservative and a decision was made to accept over-provisioning of 
CPU/RAM since GPUs are the most expensive and scarce resource; we have observed that users who hire GPUs for DL jobs 
spend a lot of money and expect the best performance. We have sometimes observed users using less CPU/RAM
than the recommended T-shirt sizes, but rarely more. Any resulting fragmentation in the cluster has not
been a problem, because there are always a small percentage of CPU-only DL jobs which can be flexibly 
scheduled to avoid resource wastage.}

\subsection{Scale test of \DLaaS cluster}
Prior to production, exhaustive testing was conducted on \DLaaS to ensure its performance and dependability under different load scenarios. 
The cluster under test has about 680 GPUs. 
In particular, scale tests were conducted under a light-load (LL) and a heavy-load (HL) scenario. 
In the light-load scenario, a total of 70 concurrent DL training jobs were executed while in the heavy-load scenario 
there were 700 concurrent DL jobs. The choice of 700 concurrent jobs ensured that we are stressing all the GPU resources and also testing queuing. 
In each scenario, there was a mix of jobs running on different GPU types. Table~\ref{tab:loadscenario} 
shows the number of jobs running on different GPU types in these scenarios. The jobs in each scenario had a staggered start in four batches; K80 jobs started in two batches in the first 15 mins, P100 jobs batch started after 30 mins, and V100 jobs batch after 32 mins. Each job is a Resnet-50/TensorFlow-1.5 training job using Imagenet1K dataset ($\sim$1.3M images), with dataset stored on \IBM Object Storage Service and  mounted using s3fs on each learner. With 700 concurrent jobs the cluster achieved on average aggregate throughput of $\sim$837 iterations/sec or 54K images processed/sec. While none of the jobs failed in the ligh-load scenario, 12/700 jobs were stuck in the heavy-load scenario and resulted in stale
learner pods. Further investigation revealed that these were deployed on nodes with hardware failures and all these nodes were later cordoned. Once the nodes were cordoned, Kubernetes automatically restarted the DL jobs and executed them to completion.
No failures related to \DLaaS software stack were observed.

\begin{table}
    \centering
    \begin{tabular}{|c|c|c|c|}
    \hline
        GPU-type-batch\# & jobs-LL & jobs-HL & start time \\
        \hline
        K80-batch1 &   30 & 300 & first 1 min \\
        K80-batch2 &  24 & 240 & after 15 min \\
        P100-batch3 & 11 & 110 & after 30 min \\
        V100-batch4 & 5 & 50 & after 32 min \\
        \hline
    \end{tabular}
    \caption{Light-load (LL) and heavy-load (HL) job mix}
    \label{tab:loadscenario}
\end{table}

\begin{figure}
    \centering
    \includegraphics[width=0.9\columnwidth]{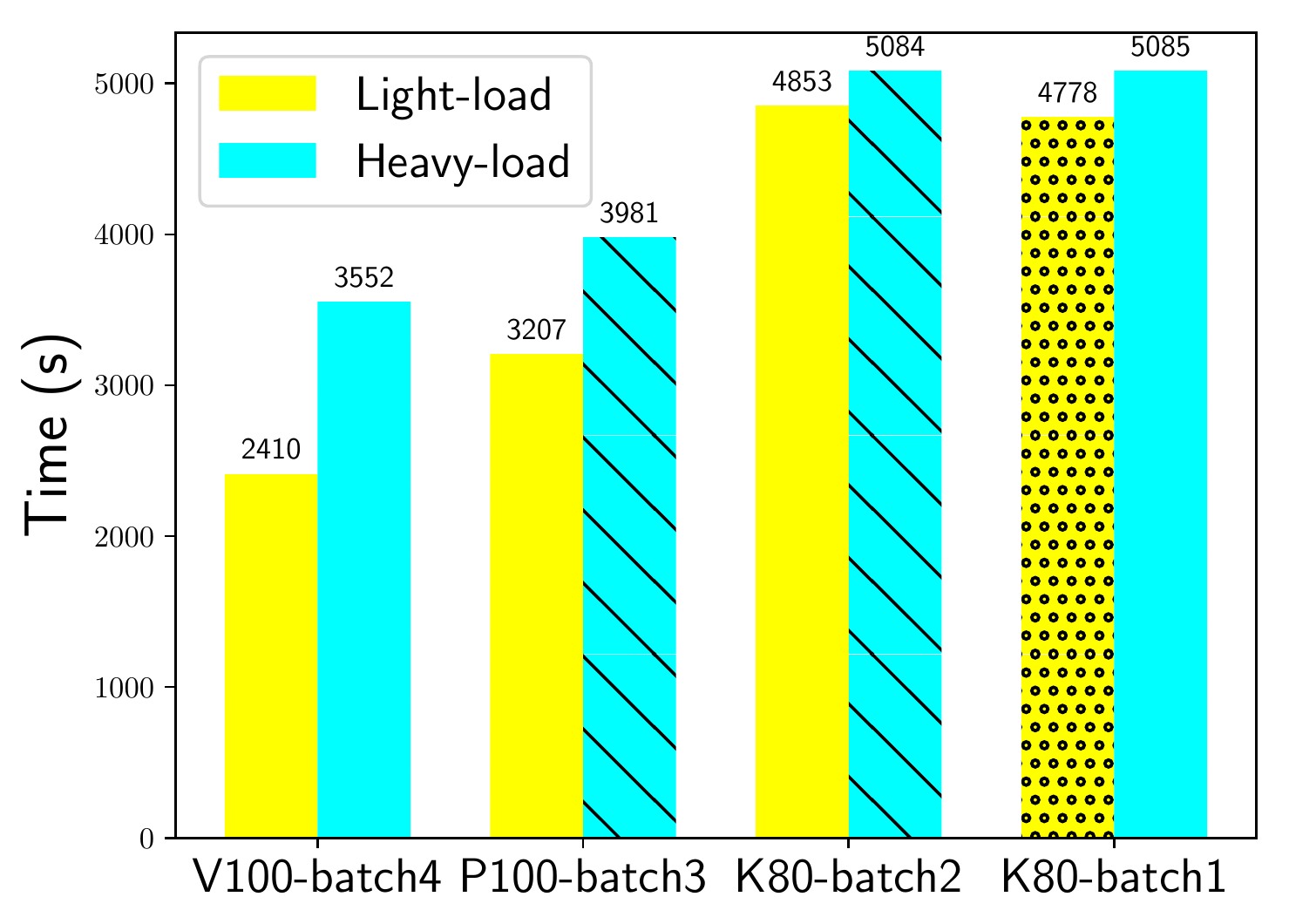}
    \caption{E2E job runtime by GPU-type for light-load and heavy-load scenarios}
    \label{fig:e2eTime}
\end{figure}

Figure~\ref{fig:e2eTime} compares the average run time of jobs on different GPU types for the two load scenarios. Observe that K80 jobs have the lowest performance degradation (6-8\%) at heavy-load compared to low-load, followed by P100 (24\%) and then V100 (51\%). This is due to the staggered start of jobs; by the time V100 jobs are running, the load is at its peak, and hence the shared resources (network and cloud object storage bandwidth) start impacting performance.

\subsection{Failure Analysis}

\begin{table*}[htb]
    \centering
    \begin{tabular}{|c|c|c|}
    \hline
        failure reason & message & \% of pods \\
        \hline
        Binding Rejected & Operation cannot be fulfilled on pods/binding "$pod\_name$":  & 17.05 \\
         &  pod $pod\_name$ is being deleted, cannot be assigned to a host &  \\
         \cline{2-3}
         & Timeout: request did not complete within allowed duration:  & 0.169 \\
        \cline{2-3}
         & pods "$pod\_name$" not found  & 1.603 \\
        \hline
        Assume Pod failed & pod $pod\_name$ state wasn't initial but get assumed  & 0.169 \\
        \hline
        persistentvolumeclaim & persistentvolumeclaim "$volumn\_name$" not found (repeated \# times) & 1.94\\
        \hline
        skip deleting pods & skip schedule deleting pod: $pod\_name$ & 15.1 \\
        \hline
        No nodes available & No nodes are available that match all of the predicates: $predicates$ & 64.0 \\
        \hline
    \end{tabular}
    \caption{The log messages indicating the reasons of scheduling failures in the cluster}
    \label{tab:scheduling_failure_reasons}
\end{table*}

\begin{figure}[htb]
    \centering
    \includegraphics[width=\columnwidth]{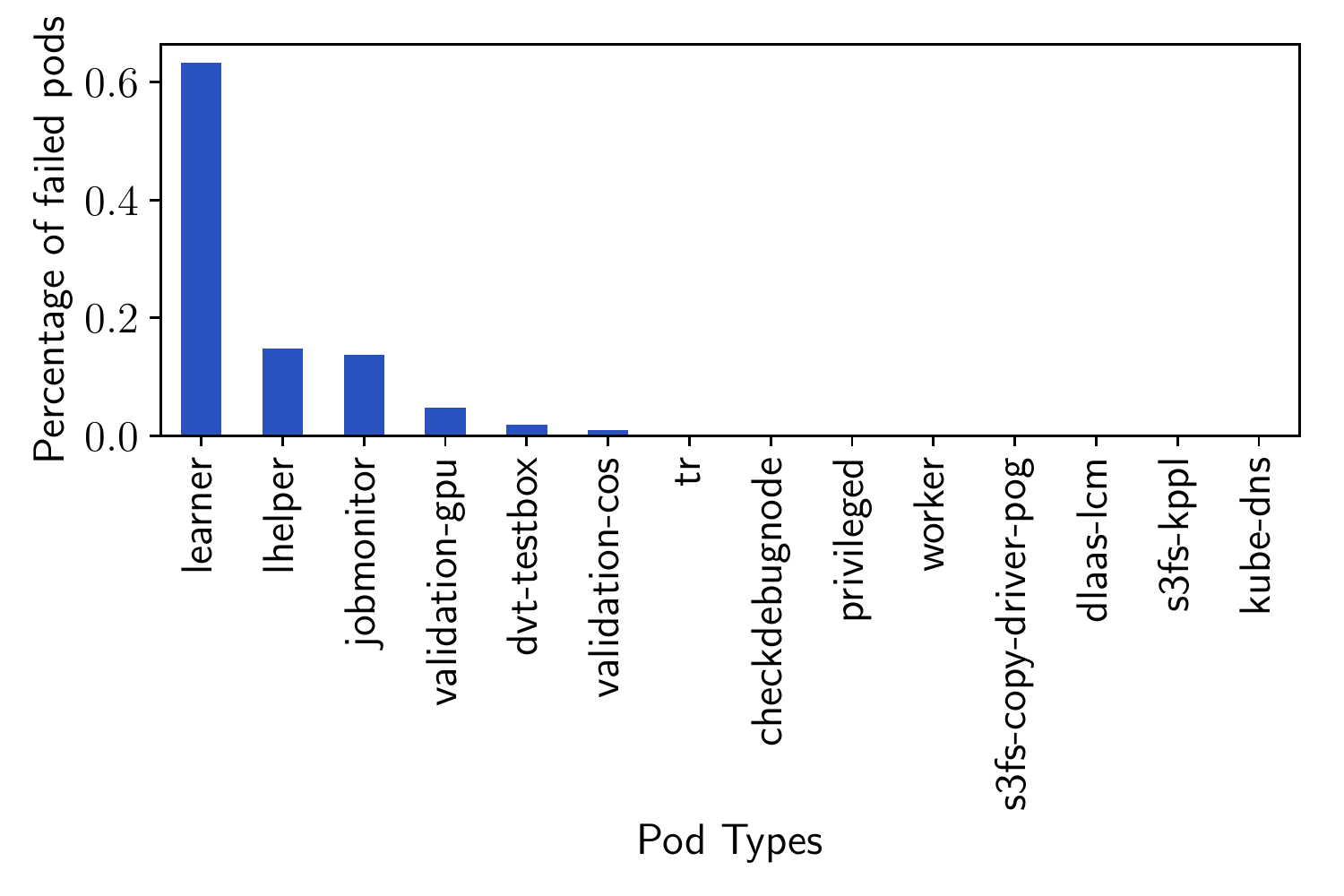}
    \caption{Distribution of scheduling failures over pod types}
    \label{fig:scheduling_failure_distribution}
\end{figure}

\begin{figure}[htb]
    \centering
    \includegraphics[width=0.9\columnwidth]{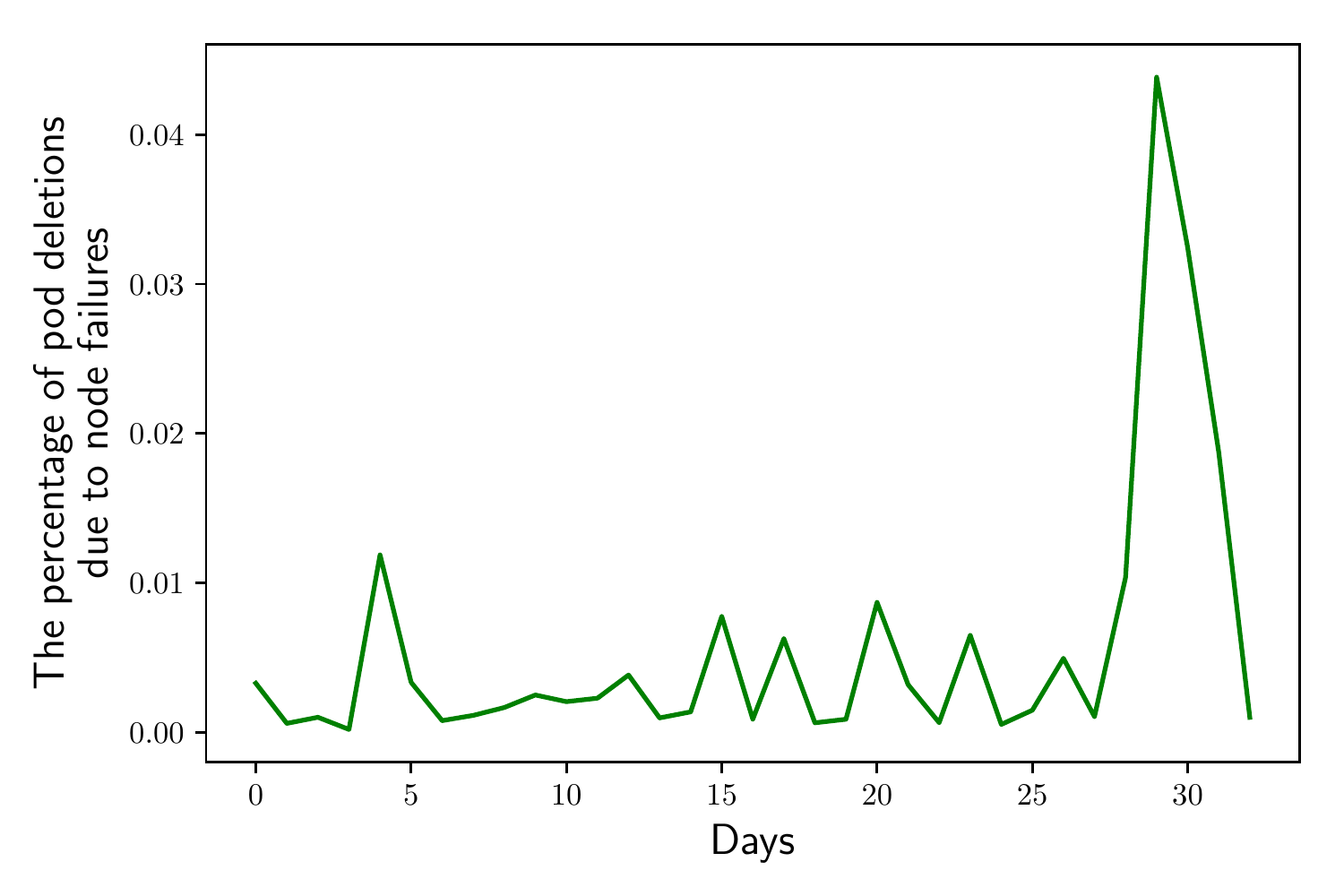}
    \caption{Percentage of pod deletions due to node failures}
    \label{fig:pod_deletion_percent_over_time}
\end{figure}

\begin{figure}[htb]
    \centering
    \includegraphics[width=0.9\columnwidth]{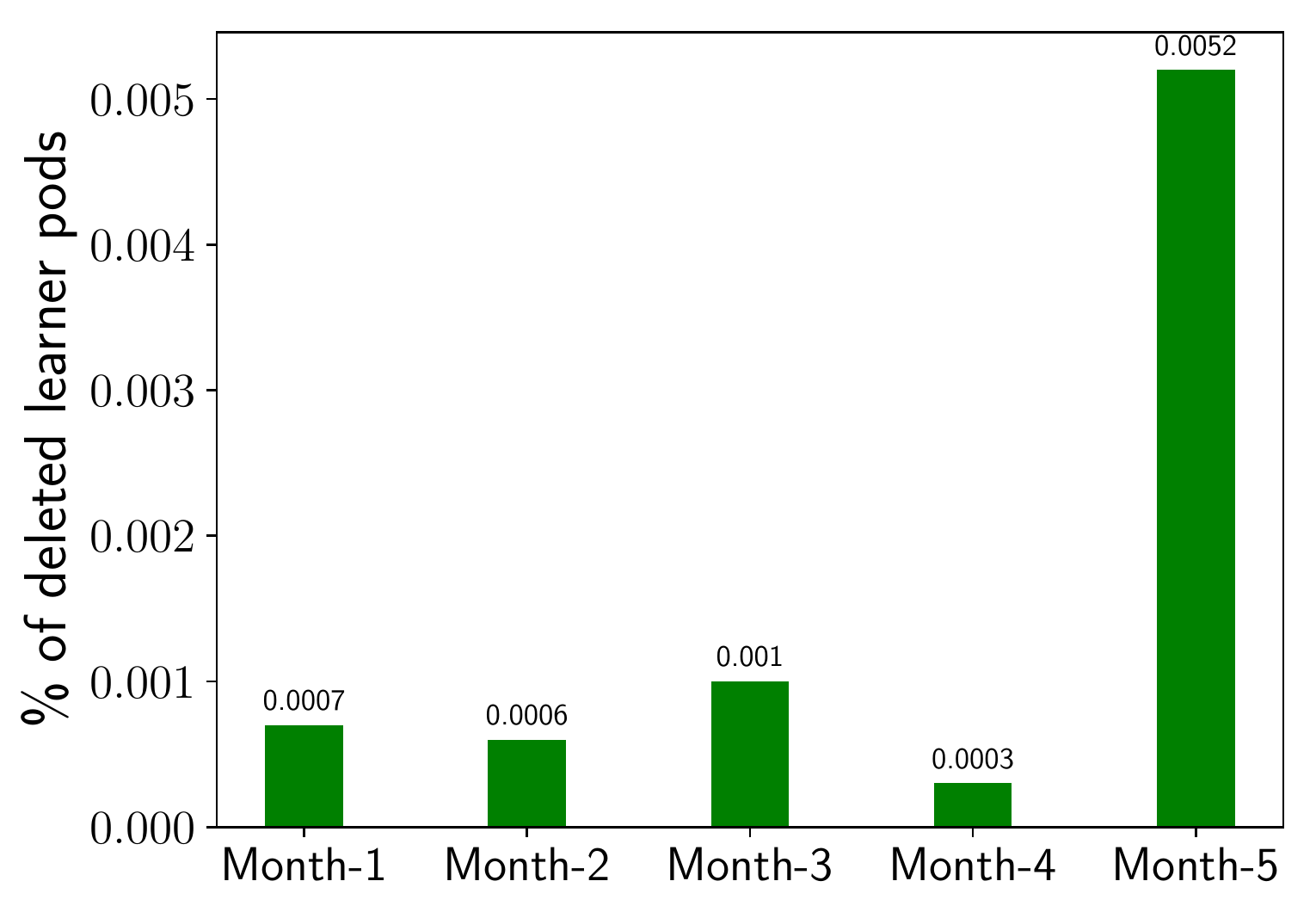}
    \caption{Percentage of \emph{learner} pod deletions due to node failures}
    \label{fig:learner_pod_deletion_percent}
\end{figure}

One typical type of failures observed is pod scheduling failure. It happens when Kubernetes scheduler fails to place a pod on any nodes in the cluster.
In a job that consists of multiple pods, failing to place one of the pods can result in the whole job pending in the cluster, taking a lot of resources by those successfully placed, and rescheduling the failed scheduling pod repeatedly.
We first get the unique pod names from all Kubernetes scheduler logs with \texttt{FailedScheduling} keyword in a 4 month period on a cluster
with 680 GPUs. We find that the distribution of scheduling failures across different types of pods is not uniform. 
As shown in Figure \ref{fig:scheduling_failure_distribution}, among all types of pods running in the cluster, only 14 of those had ever failed to be scheduled.  More than 60\% of failed scheduling pods are \textit{learners}, followed by around 15\% of failed \textit{lhelper} scheduling pods.

We then parse the messages of all \texttt{FailedScheduling} logs to understand the main reasons of the scheduling failures, as shown in Table ~\ref{tab:scheduling_failure_reasons}. 
Around 64\% of pods failed to be scheduled due to \texttt{No nodes are available that match all of the predicates}.
In Kubernetes scheduler, if one condition requested by the pod cannot be met on a certain set of nodes, these nodes are ``predicated'', namely the pod cannot be placed on these nodes.
By analyzing the predicates, we find the top ones are \\ \texttt{Insufficient alpha.kubernetes.io/nvidia-gpu},\\ \texttt{MatchNodeSelector} and \texttt{NodeUnschedulable}, indicating that these pods failed to be scheduled because all nodes in the cluster were either running out of GPU resources, or did not have the GPU framework requested, or were not in ready status to run pods.
The above predicates and the failure distribution indicate that 1) the majority of scheduling failures were caused by certain insufficient resources in the cluster; 2) such scheduling failures would only impact the pods (\textit{learner}) who requested the scarce resources.

The other type of failures are due to worker node failures. 
From the Kubernetes \textit{controller-manager} logs, we observed logs that recorded the nodes' health status.  We noticed that when worker nodes became \texttt{NotReady}, the \\ \texttt{NodeControllerEviction} component in Kubernetes would delete all pods running on the worker.
Worker node can become \texttt{NotReady} inevitably due to various reasons, such as hardware failures, OS updates, networking container daemon failure, etc. 
For pods that are necessary for running training jobs, such as learner pods, the whole job needs to be canceled and restarted if one of these pods are deleted due to worker failures.  All GPU resources that have been consumed for training the job are wasted.
In Figure \ref{fig:pod_deletion_percent_over_time}, we show that overall the percentage of pod deletions due to node failures is within 5\% over time.

We also show the impact of node failures on training jobs. In Figure \ref{fig:learner_pod_deletion_percent}, we show the percentage of \textit{learner} pods being deleted due to node failures in every month. Assuming all failed \textit{learner} pods belonged to different training jobs, we see that the cancellation of jobs due to the deletion pods was below 1\% in the months we studied.

\section{Related Work}
\label{sec:related}

A general introduction to fault tolerance in distributed systems is available in \cite{Koren:2007:FS:1206164}; \cite{HASAN2018156,Sharma:2016:REE:3005778.3005795} provide a detailed categorization of failure types grouped into proactive and reactive approaches. \cite{Cheraghlou2016ASO} summarizes fault tolerance architectures. One can distinguish reactive approaches, i.e. those in response to a failure, from proactive ones, i.e. such predicting failures and taking proactive action, e.g. by preemptive job migration, cmp. \cite{Wang:2008:PPL:1413370.1413414}. As outlined in \cite{Sharma:2016:REE:3005778.3005795}, the most relevant reactive approaches include checkpointing (coordinated, uncoordinated or communication induced), replication and logging, all of which are employed by \DLaaS\ . Sampaio et al. \cite{DBLP:conf/isami/SampaioB17} showed that proactive fault tolerance can be superior to double replication both in the number of successful executions as well as energy efficiency.

Regarding bigger clusters, \cite{Egwutuoha2013} summarizes reliability in HPC settings. Dongarra et al. \cite{doi:10.1177/1094342010391989} estimated that, as we approach exascale computing, the number of CPUs, disks and memory will increase exponentially and thus the mean time between failures (MTBF) will converge towards just 1 min. Besides the high frequency of failures, there is also the large variety of errors to deal with -- general software and hardware failures, human error, network issues, power outages, cyber attacks, etc. Furthermore, they point out that new technologies like phase-change RAM affect resiliency both positively (e.g. faster checkpointing) as well as negatively (e.g. difficulty of persisting accelerator state; Intel is reported to expect a power consumption increase by 15-20\% due to correctness checks). It should also be noted that acquisition cost pressure increasingly leads to utilization of commercial off-the-shelf (COTS) components, which in turn exacerbates failure rates and thus emphasizes the importance of dependability. In this context, chaos engineering, e.g. implemented in Netflix' Simian Army as well as Facebook Storm, has become popular. Chaos Engineering deliberately injects failures into production systems, enabling engineers to test fault tolerance. It is described in more detail by Gunawi et al. \cite{faas}.

However, less attention has been paid to DL workloads in aforementioned research; fault tolerance for DL workloads are very relevant to products like like IBM Watson Machine Learning~\cite{ibm-wml}, Amazon SageMaker~\cite{amazon-sagemaker}, Google Cloud Machine Learning~\cite{google-cloud}, and Microsoft Azure~\cite{microsoft-ml}. DL frameworks like TensorFlow, PyTorch and MXNet are largely limited to user-controlled checkpointing, but generally leave rescheduling of failed pods to the cluster manager, even though monitored sessions can at least facilitate session recovery from a checkpoint. A more detailed exploration of their capabilities can be found in \cite{Georgia2016SoftwareFF} and \cite{8038464}. 

Li et al. \cite{Li:2014:SDM:2685048.2685095} claim that their parameter server approach provides recovery from non-catastrophic machine failures
within 1s, without interrupting computation and thus continuous fault tolerance. The systems they compare against, GraphLab~\cite{graphlab}, REEF~\cite{reef} and Naiad~\cite{naiad} only provide checkpointing. MLBase~\cite{mlbase} and Spark~\cite{spark} provide ``resilient distributed datasets-based'' fault tolerance, i.e., the source dataset and the list of transformations on it are stored reliably, and the transformations
are recomputed in the event of failure. Petuum~\cite{conf/kdd/XingHDKWLZXKY15} claims that fault tolerance for up to 100s of machines is implemented via checkpoint-and-restart, whereas scaling it to 1000s of machines is listed as future work. Parameter servers have found wide adoption and are also supported by \DLaaS\ . \cite{Amatya2017WhatDF} discusses several approaches to fault tolerant deep learning with MPI and claims that previous approaches are either not fault tolerant or not HPC-ready, whereas their FT-Caffe proposal is claimed to fulfil both requirements simultaneously. \DLaaS\  uses MPI through its Horovod integration.

Unfortunately, neither Google's TFX \cite{Baylor:2017:TTP:3097983.3098021} nor Uber's Michelangelo \cite{Michelangelo} or Facebook's FBLearner \cite{FBLearner} have elaborated too much on their approaches towards fault tolerance which hopefully increases the usefulness of this paper.
The Hemingway tool~\cite{pan:17} guides the selection of appropriate algorithms and cluster size for distributed
training jobs by using predictive models for job runtime and convergence~\cite{venkataraman:16}. However, these models
do not account for resource contention among jobs which typical in a cloud environment. The SLAQ framework~\cite{zhang:17} explores
quality-runtime tradeoffs across multiple jobs to maximize system-wide quality by adjusting resource allocations of all
running jobs. 
There is little relevant work on efficient scheduling of DL jobs on heterogeneous platforms~\cite{Peng:2018,Park:2018,222611}.
The scheduler in \cite{Peng:2018} improves the turn around time of DL jobs in a cluster by using predictive modeling to estimate the number of training 
steps and the impact of different hardware resources on the training speed. 
Information about distribution of job runtime from historical data is used by the scheduler in ~\cite{Park:2018}  to effectively pack jobs with diverse time and placement concerns while GPU time-slicing is proposed in ~\cite{222611} to improve latency and efficiency of training deep learning models in cluster environment.

\section{Conclusion}~\label{sec:conclusions}

This paper presents \DLaaS, a platform to support the distributed training of deep learning models in the cloud. Data scientists and cloud service developers want a platform that makes it easy to use existing DL frameworks, but support the dependability, scalability, efficiency, multi-tenancy, and security properties people have come to expect from cloud services.
A careful selection of the lowest common denominator features across frameworks, such as filesystem support, configuration via environment variables, process exit codes, and logging to standard out were used to build a DL framework agnostic middleware layer. The Kubernetes scheduling algorithms had no concept of atomic job deployment, and we incorporated the BSA gang scheduling algorithm to avoid scheduling deadlocks. Similarly, optimized storage drivers allowed us to support cloud storage services for DL training data while keeping DL frameworks unaware of any cloud services. A number of cloud services were used in the design, including an object storage service for training data, etcd for distributed coordination, MongoDB for metadata storage. A careful coordination of health checkers and failure detectors at the hardware, cluster manager, middleware, and DL frameworks is used to orchestrate the provisioning of all the resources needed to deploy a DL job in a fault-tolerant manner.

We have operated \DLaaS within \IBM for over a year and the system has maintained high availability with low failure rates in an environment where nodes fail or are removed for maintenance, and new resources added at any time. Most failures are due to faulty nodes or cluster manager provisioning errors, which can be recovered by restarting jobs either from the beginning or from a checkpoint. Experiments demonstrate that the system scales to large numbers of concurrent jobs, but we observed that running 700 concurrent jobs in a 680 GPU cluster resulted in performance degradation. This was mainly due to network capacity and storage throughput limits, and not an inherent limit of \DLaaS itself. Empirically, the service has scaled to support an increasing number of users and without any known security compromises.

We have open-sourced major portions of this platform (\url{https://github.com/IBM/FfDL}); we plan to add cluster traces and other artifacts there, and hope it can be a foundation for further research in this area.

\begin{acks}
We would like to thank the anonymous reviewers of Middleware'19 and our shepherd Derek Murray for their insightful feedback.
\end{acks}

\bibliography{dlaasdep}

\end{document}